# RS-Coded Adaptive Dynamic Network for Reliable Long-Term Information Transmission in Disturbed Multimode Fiber

Yang Hu[1], Minyu Fan[1], Kun Liu[1], Songsong Zhu[2], Nan Jiang[2], Sha Wang[1], *

1. College of Electronics and Information Engineering, Sichuan University, Chengdu, 610064 China

2. State Key Laboratory of Oral Diseases & National Center for Stomatology & National Clinical Research Center for Oral Diseases & Department of Orthognathic and TMJ Surgery, West China Hospital of Stomatology, Sichuan University, Chengdu 610041 Sichuan, China

*shawang@scu.edu.cn

Multimode fiber (MMF), due to its large core diameter and high mode capacity, holds potential in high-speed communications. However, inherent modal dispersion causes output speckle distortion, and transmission characteristics are sensitive to environmental disturbances, limiting its reliable application. Conventional transmission matrix (TM) methods face challenges such as complex calibration and environmental sensitivity. Although current deep learning approaches demonstrate reconstruction potential, they struggle to overcome error accumulation caused by fiber mode drift and lack sufficient environmental adaptability. To address this, the present study proposes an adaptive transmission framework named Residual Reed-Solomon Dynamic Network (RRSDN), which integrates Reed-Solomon (RS) error correction coding with deep residual learning forming a closed-loop system that jointly optimizes encoding, transmission, and reconstruction, to tackle the key challenges of mode instability and error accumulation in dynamic scattering channels. Experimentally, high-fidelity real-time transmission of a 16 × 16 pixel video stream (H.265 compressed) with zero frame loss and 100% symbol accuracy was achieved under conditions of a 100-meter MMF with manually applied disturbances and no temperature control. This work proposes a solution for stable optical transmission in complex channels. Plus, it integrates error correction coding with neural network training, laying the foundation for adaptive optical systems in longer-distance and more complex scenarios.

**Introduction：**

Optical fiber information transmission, as a core technology in modern information transmission, is based on the guided-wave propagation of light within the fiber medium. This technology features wide bandwidth, low loss, strong interference resistance, compact structure, and low cost, making it indispensable in high-speed, large-capacity information transmission. Compared to single-mode fiber (SMF), multimode fiber (MMF), with its larger core diameter and numerical aperture, can support parallel transmission of hundreds of spatial modes[1]. This structural characteristic enables MMF to achieve higher mode coupling efficiency and optical power handling capacity in information transmission[2-3]. However, its modal dispersion differs by orders of magnitude from that of single-mode fiber[4], causing severe intermodal delay during transmission. This results in indistinguishable speckle patterns at the output[5], preventing direct recovery of input information.

Current mainstream MMF transmission technology is based on the transmission matrix (TM) theoretical framework[6-8]. The TM, as a mathematical tool describing the linear transformation between input and output optical fields, enables extraction of fiber eigenmode information through singular value decomposition (SVD), thereby characterizing the channel properties. By computing and fitting the TM and modulating the input optical field accordingly, the desired optical field distribution can be obtained at the fiber output, thus suppressing mode mixing and intermodal interference in MMF. In 2015, Martin Plöschner and colleagues demonstrated that MMF systems are predictable[6]. By constructing the TM and applying SVD analysis, the authors decoded the fiber's cylindrical symmetry and achieved high-fidelity prediction of the output optical field. They advanced transmission matrix technology from theory to practice, providing a physical foundation for TM-based output optical field control. In 2018, Moussa N'Gom and colleagues proposed a reference-arm-free TM measurement method[7], using only an imaging system and phase modulator, simplifying the conventional interferometric measurement setup. The TM was directly retrieved from output light intensity using a semidefinite programming (SDP) algorithm, avoiding complex interferometric setups. This overcomes the reliance on reference light in TM measurement and promotes the application of TM technology in portable devices[8]. However, TM-based methods have inherent limitations: changes in MMF transmission modes caused by environmental disturbances lead to rapid TM invalidation[9-11], requiring frequent and time-consuming re-measurement[12] or construction of large matrix libraries[13], which severely restricts their application in dynamic environments.

With advances in deep learning for solving physical inverse problems[14], data-driven end-to-end mapping methods[15] offer new approaches for MMF transmission, demonstrating significant potential in addressing inherent instability and environmental sensitivity[16]. Early studies focused on using deep learning to decode and reconstruct speckle patterns at the output of MMF[17-18] in order to recover the input information. In 2018, Babak Rahmani and colleagues first demonstrated the application of deep learning in MMF imaging[19]. Using supervised learning, the neural network was trained to recognize and reconstruct distorted speckle patterns at the fiber output, successfully recovering the intensity or phase information of simple images such as handwritten digits and letters. In the same year, Shachar Resisi and colleagues trained a network to learn the mapping between input images and output speckle patterns[20], overcoming the limitation of real-time calibration required by conventional TM methods. They were the first to demonstrate that deep learning can address signal mixing caused by mechanical disturbances in fibers, using a convolutional neural network to reconstruct input images from speckle patterns output by a 1.5-meter MMF. These early studies established direct mappings between input and output, reducing reliance on precise modeling of the fiber channel, and can be regarded as an initial step toward relaxing the stability requirements of fiber transmission through neural networks.

However, the signal transmission in MMFs is highly susceptible to external physical factors such as bending and temperature variations, leading to dynamic changes in propagation modes. Conventional single training datasets cannot cover the full range of environmental disturbances, resulting in degraded performance of pre-trained models in real dynamic environments. To address this issue, studies have explored joint training (or multi-scenario training) to enhance the model's adaptability to a wider range of disturbances[21]. By collecting data under varying environmental conditions and performing joint training, the model can learn more robust feature representations, thereby achieving a certain level of resilience to unseen environmental variations. For example, incorporating training data with varying degrees of bending and temperature fluctuations can enhance the model's generalization in practical applications and improve its adaptability to a broader range of disturbances. In 2023, Shuqi Zhang and colleagues proposed a pilot-assisted joint training framework to address the dynamic mode coupling in MMF caused by mechanical bending and temperature drift during practical deployment[22]. They collected transmission data under various bending states and temperature conditions in a 100-meter commercial MMF to construct a multi-disturbance scenario dataset. The disturbance-invariant features learned through joint training enable the model to generalize to unseen bending angles and temperature fluctuation scenarios.

Although joint training can extend the model's adaptability to a range of disturbances, more flexible mechanisms are still required to handle real-time and unpredictable environmental changes[23]. The introduction of self-supervised learning paradigms offers new approaches for real-time adaptation to dynamic measurement changes. By leveraging the intrinsic structure of the data as a supervisory signal[24], the model can continuously learn and update without external labels[25]. In 2024, Ziwei Li and colleagues proposed a dynamic memory framework that uses a mixture of experts model to adaptively and accurately track changes in light propagation patterns within MMFs[26], achieving efficient transmission of high-resolution video. The framework employs a self-supervised mechanism, enabling the model to adjust its internal state based on real-time observations, thereby dynamically adapting to channel variations. However, this self-supervised mechanism still struggles to fully avoid the risk of error accumulation: when the network predictions deviate, the resulting pseudo-labels introduce errors that mislead model updates, potentially causing the model to fall into sub-optimal states or erroneous convergence.

To address the key challenges of mode instability and error accumulation in dynamic scattering channels, this study proposes an adaptive transmission framework Residual Reed-Solomon Dynamic Network (RRSDN) based on a residual network structure[27] collaboratively optimized with Reed-Solomon (RS) coding[28]. The well-established RS error-correcting code from information theory is integrated into the ResNet-based adaptive learning process. The RS encoder introduces redundancy at the transmitter, and the RS decoder at the receiver can detect and correct byte-level errors occurring during transmission and reconstruction in real time[29-30]. This blocks erroneous labels from entering the self-supervised training loop at the source, effectively suppressing error accumulation. Meanwhile, the ResNet architecture efficiently captures both transient and slowly varying channel dynamics[31]. The prior constraint mechanism of RS code symbols enables RRSDN to balance training efficiency and prediction accuracy, thereby dynamically adapting to drastic changes in MMF transmission patterns. Under the condition with no temperature control and allowing free lateral displacement of the fiber, transmission tests with 8×8 and 16×16 pixel information were conducted on a 100-meter MMF. Results show that the system maintains stable prediction accuracy for binary amplitude modulation patterns and successfully achieves high-fidelity reconstruction of MMF video streams. This work integrates RS error-correcting coding with deep residual learning, optimizing the compatibility between discrete coding and continuous neural network training, and provides a solution for MMF transmission systems that combines theoretical advancement with engineering practicality. The cross-domain fusion characteristic of this framework opens new directions for adaptive optical information transmission in complex media channels and offers a new paradigm for the seamless integration of discrete coding operations with continuous neural network training.

**Results：**

**Dynamic Reconstruction and Error Correction**

Considering the strong burst error correction capability of RS coding and its suitability for byte errors, this study constructs an RS-coded MMF information transmission system. The core process, shown in Fig. 1(a), includes three main modules: encoding and modulation, optical field transmission, and intelligent reconstruction. During the encoding and modulation process, the information to be transmitted is first converted into an 8-bit binary byte sequence, which is then grouped according to a preset length by the RS encoder to generate a parity code sequence with adjustable error correction capability. The information and parity codes are integrated according to a specific structure, with the encoding process following the systematic code standard, as shown in Fig. 1(b). A binary amplitude pattern compatible with digital micromirror device (DMD) spatial modulation is generated, as shown in Fig. 1(c). The pattern is dynamically displayed by the DMD to spatially modulate the input optical field, and the modulated optical signal is coupled into the MMF transmission channel.

During fiber transmission, mode coupling effects and intermodal dispersion cause the output to form a randomly distributed speckle field. The characteristics of the speckle field are closely linked to the input pattern and channel state. They are captured by a scientific-grade CCD (Basler acA640-750um) at 100 frames per second, forming the dataset for subsequent reconstruction network training and validation.

The reconstruction network adopts a three-level subnetwork architecture as shown in Fig. 1(d): short-term subnetworks S1 and S2 are alternately rebuilt every five time windows to handle transient disturbances such as mechanical vibrations and sudden temperature shifts. The long-term subnetwork L freezes its parameters after self-pretraining and is used to model the slowly varying channel dynamics. Fig. 1(e) illustrates the subnetwork structure. Each subnetwork is built upon a ResNet residual block and consists of dual $3 \times 3$ convolutional layers in series (with 8 channels), followed by batch normalization and dropout layers ($p = 0.4$). ReLU is used as the intermediate activation function, and Sigmoid is applied at the output layer. The hierarchical fusion module dynamically integrates the three output features using an uncertainty-weighted mechanism. Detailed procedures are provided in the supplementary.

The reconstruction process follows a two-stage optimization strategy, as illustrated in Fig. 1(f). In the pre-training stage, an accurately registered pattern-speckle dataset is used to establish an initial mapping model via cross-entropy loss. The model is then transferred to a self-supervised stage, where speckle intensity data from the MMF output is collected in real time and processed in batches within temporal windows. Gated recurrent units reset the short-term memory states of sub-networks S1 and S2. An embedded RS decoder performs dynamic error correction. Using the corrected signals, loss is computed and network parameters are updated via direct gradient propagation. Effectively suppresses error accumulation caused by isomorphic variations in MMF transmission modes, enhancing adaptability to dynamic channels.

The RS decoding module integrated at the output layer operates over the Galois Field ($GF(2^8)$) to locate and correct symbol-level (byte) errors within the predefined error correction capacity $t$. Detailed procedures are provided in the supplementary material. The corrected outputs are fed back through a gradient passthrough mechanism to update the network parameters, forming a closed-loop learning process of "prediction–correction–optimization."

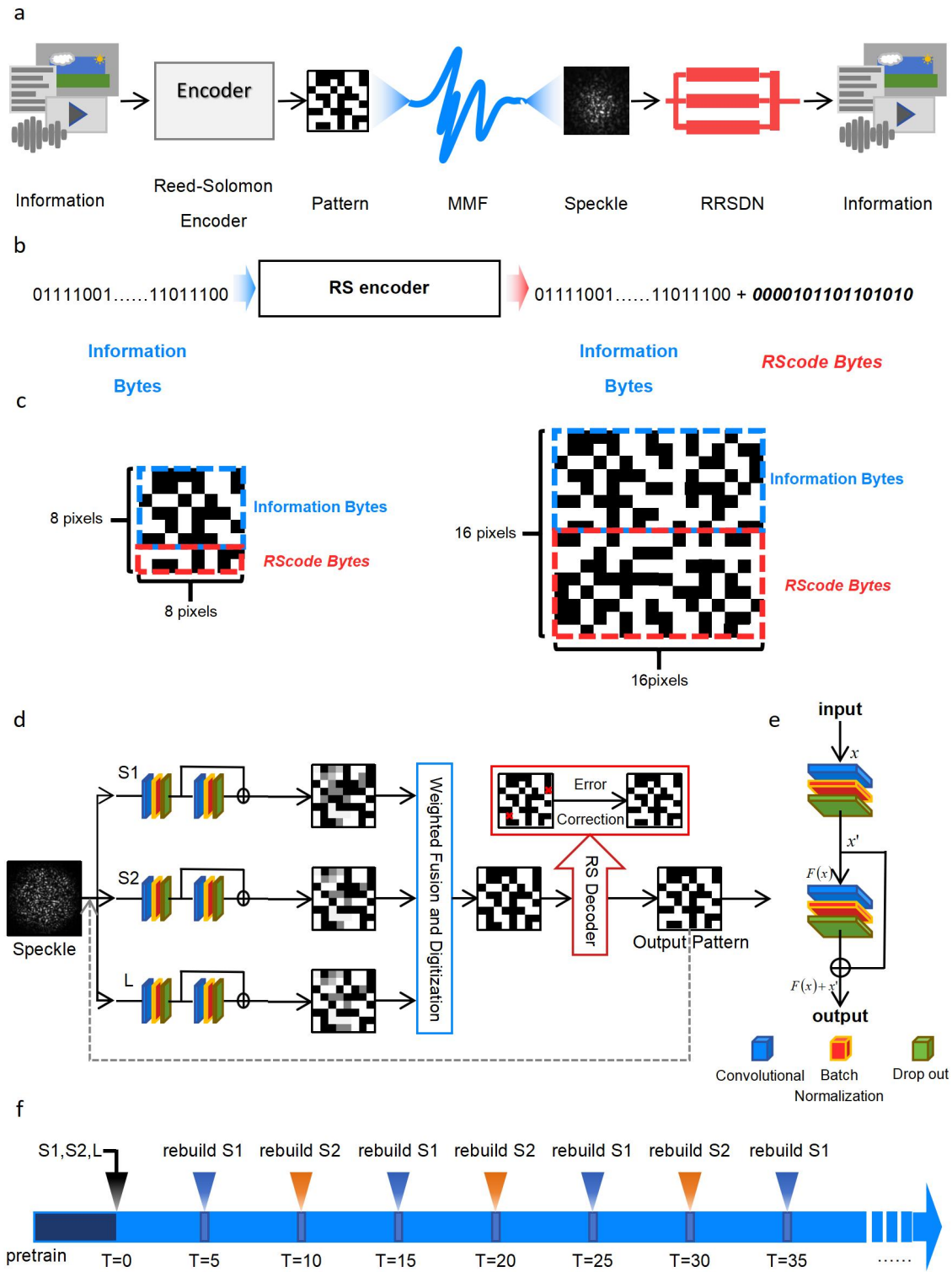

**Fig.1:Principle of RRSDN transmission in disturbed MMF. a** The core workflow of RS-coded MMF information transmission;**b** RS encoder generates RScode Bytes according to information Bytes and predetermined data length;**c** Structure of 8×8 and 16×16 mask that will be displayed on DMD;**d,e** The network architecture of RRSDN,where **d** the data is fed into three sub networks and processed through weighted fusion and digitization,RS decoder correct errors,then the output results are used as the next updated data,and **e** the architecture of each sub networks;**f** The pre-training and dynamic updating process of RRSDN,alternately rebuild sub networks S1 and S2 every 5 time windows.

**Information Transfer Accuracy**

Real-time MMF transmission experiments were conducted using 8×8 and 16×16 mask sizes to verify information transmission accuracy over a duration of 300 seconds. The pre-training phase lasted for the first 4 seconds and the first 20 seconds, during which speckle-pattern paired data were collected. The model was then transferred to the self-supervised reconstruction phase. Dynamic transmission prediction was performed using three architectures: RRSDN, MMDN[25], and StaticNN, which is a convolutional neural network whose parameters are frozen after pre-training . Fig. 2 compares the transmission accuracy of 8 × 8 and 16 × 16 mask size over different fiber lengths. As shown in Fig. 2(a) and (c), under low environmental disturbance, the 100-meter fiber transmission over 300 seconds exhibited a gradual decline in long-term SSIM and PCC, while short-term SSIM and PCC remained above 0.96, indicating slow drift in transmission modes. Both MMDN and RRSDN maintained high performance, with average symbol accuracy exceeding 99.5%. To evaluate the system's robustness under high disturbance, ±0.1 mm lateral random perturbations were applied manually using a displacement platform during the experiment, inducing significant changes in fiber modes. As shown in Fig. 2(b) and (d), the SSIM and PCC metrics exhibit nonlinear fluctuations, with the minimum interval frame SSIM dropping below 0.89. Under these conditions, RRSDN still demonstrates strong robustness, with an average symbol accuracy exceeding 99.99%. In contrast, MMDN is dominated by error accumulation effects: prediction errors generate false labels that trigger a negative feedback loop, causing loss function oscillations and divergence, and the prediction accuracy fails to recover. Due to the absence of a dynamic update mechanism, StaticNN's accuracy closely follows changes in fiber stability metrics, highlighting the inherent limitations of static models. During transmission with a 16×16 mask size, RRSDN achieved an average transmission accuracy improvement of over 35 percentage points compared to MMDN's 64.94%.

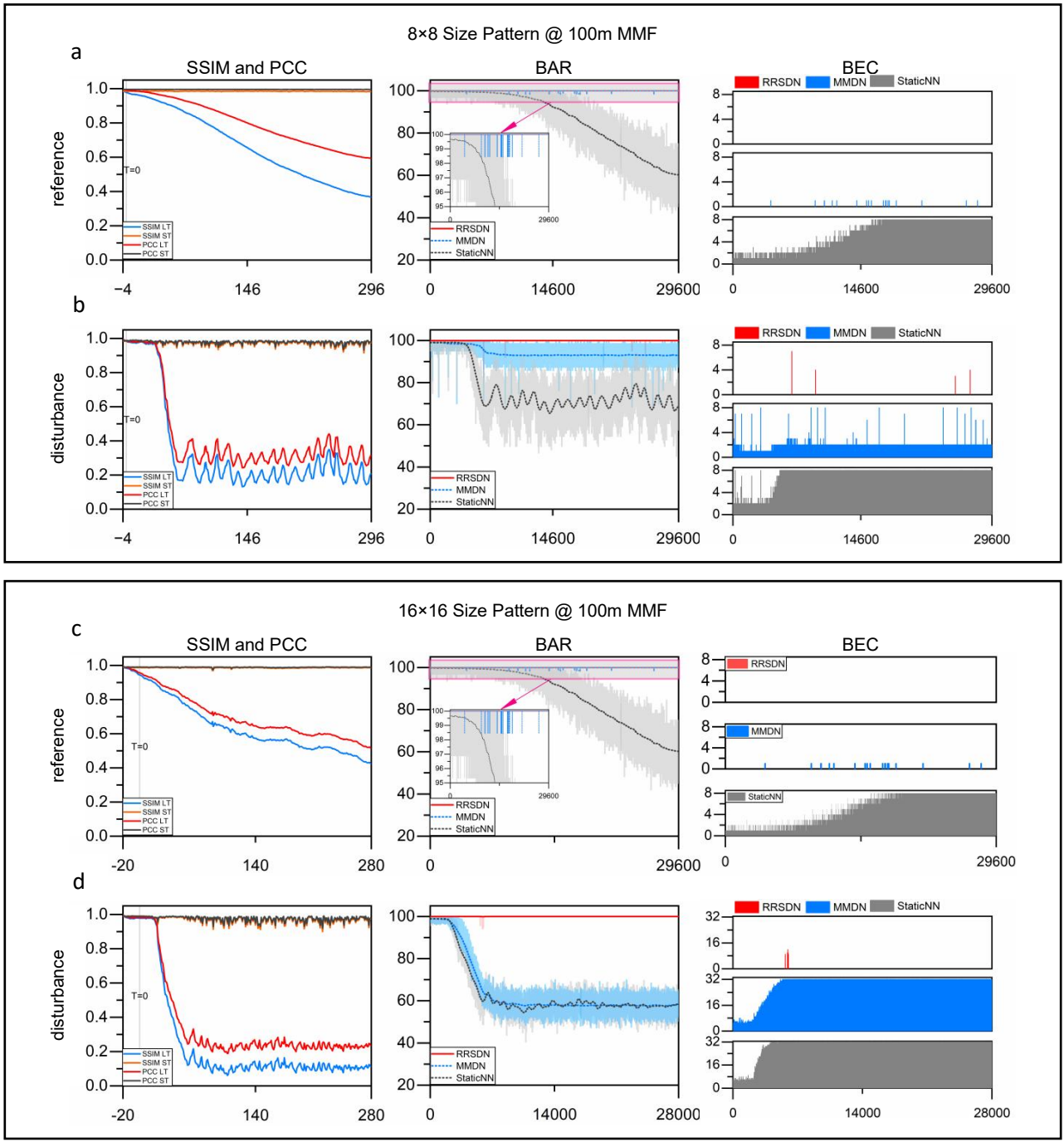

**Fig.2: SSIM and PCC of 100m length fiber and BAR, BEC values of each network. a,b** The results of 8×8 Size Pattern, where **a** no extra disturbance was introduced, as a reference, and **b** ±0.1 mm lateral random perturbations were applied; **c,d** The results of 16×16 Size Pattern, where **c** no disturbance was introduced, as a reference, and **d** ±0.1 mm lateral random perturbations were applied.

**Video Transmission Capability**

This study further conducted MMF video transmission experiments to validate the engineering applicability of the dynamic adaptive framework. The experiment employed H.265 encoding to compress the original 200×200 resolution,20 fps video stream, followed by mapping the encoded data into 16×16-bit binary amplitude masks using an RS encoder, generating a total of 28,000 mask frames. Real-time video streaming was performed through the MMF at a rate of 400 masks per second for 70 seconds. The receiver utilized the RRSDN architecture for end-to-end reconstruction. The experimental results are shown in Fig. 3. RRSDN maintained a symbol prediction accuracy of 100% throughout the transmission, successfully recovering the complete H.265 video stream without any frame loss. The reconstructed video quality was high, with the structural similarity index (SSIM) between reconstructed and original frames close to 1.0. In contrast, MMDN, lacking a dynamic error correction mechanism, exhibited a gradual decrease in prediction accuracy due to mode drift in the fiber, dropping to a minimum of 75.7%. During transmission, MMDN and StaticNN experienced high bit error rates, causing video frame data corruption. As a result, the video decoder failed to extract complete frame data, leading to severe frame loss. MMDN successfully recovered 258 video frames, while StaticNN recovered only 112 frames. Moreover, StaticNN completely failed after 18 seconds due to its inability to adapt to channel variations. These results demonstrate the essential role of the RRSDN dynamic architecture and closed-loop error correction mechanism in enabling reliable real-time video transmission.

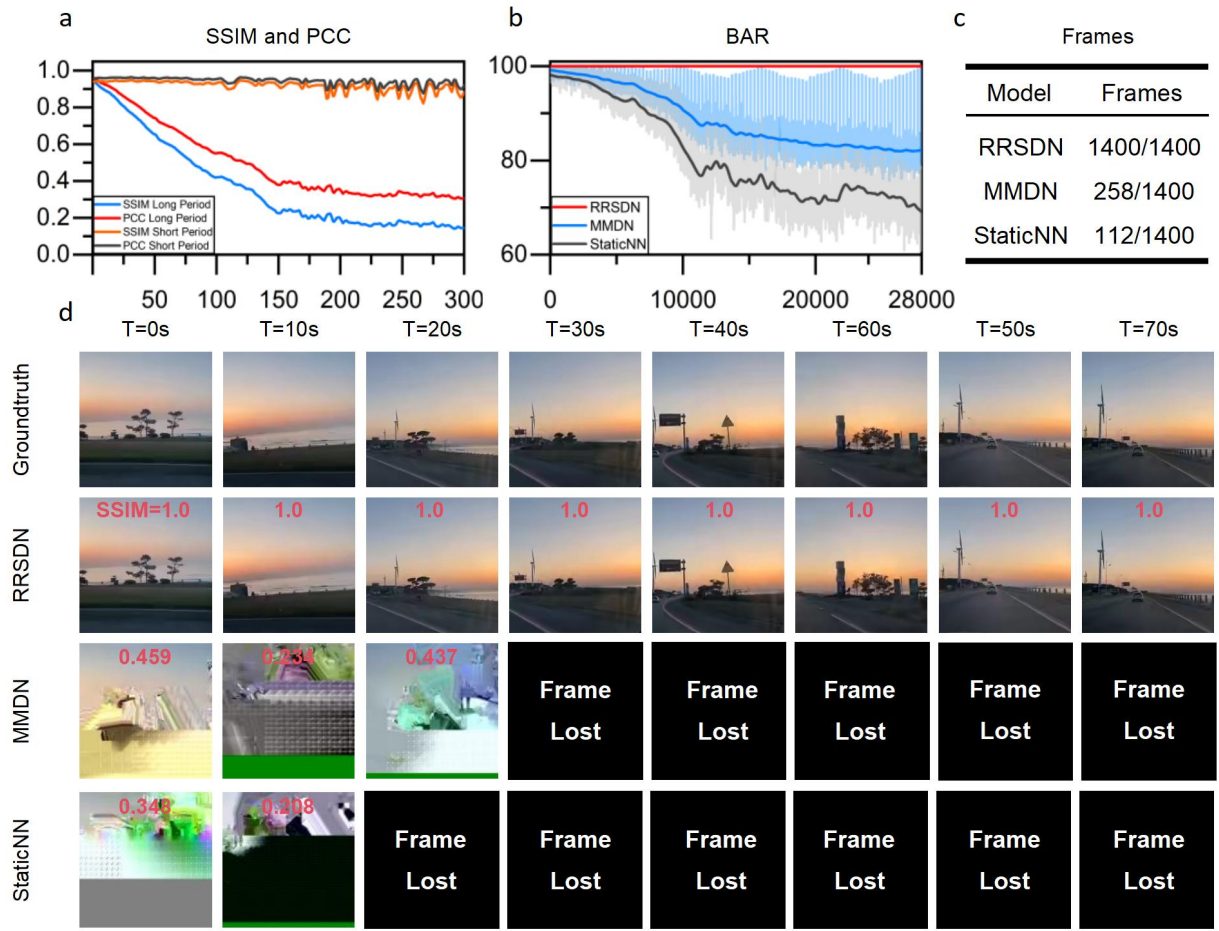

**Fig.3:Result of RRSDN on video transmission capability. a** the SSIM and PCC values of fiber during transmission;**b** the BAR values of networks;**c** The number of frames that a video decoder can decode;**d** recovery results.

**Discussion：**

This study proposes the RRSDN framework, which integrates Reed-Solomon (RS) error correction coding with residual networks to effectively address mode instability and error accumulation in dynamic scattering channels of MMFs. It offers a solution for optical information transmission over long distances and in high-disturbance environments. Experimental results show that under the condition with no temperature control and with free lateral displacement , RRSDN achieves over 99.99% symbol transmission accuracy in 100-meter MMF, improving by 35.01% compared to the MMDN method. It also demonstrates high-fidelity real-time reconstruction of video streams. The redundancy introduced by the RS encoder enables the embedded decoder to perform real-time byte-level error detection and correction based on $GF(2^8)$ operations before pseudo-label generation. This mechanism fundamentally blocks erroneous predictions from entering the training data pipeline, effectively suppressing the negative feedback loop caused by error accumulation. Experiments show that in a 100-meter fiber, compared to MMDN, RRSDN increases the proportion of effective training samples (prediction accuracy = 100%, byte errors = 0) from 10.09% to 99.95%. The system employs ResNet-based subnetworks with dual convolutional paths and hierarchical fusion to decouple transient disturbances (e.g., mechanical vibrations) from slow drifts (e.g., temperature gradients). A dynamic decoding gradient backpropagation mechanism allows decoded data to update network parameters, forming an adaptive closed-loop of "prediction-correction-optimization." With self-supervised dynamic updates, performance is further improved. RRSDN converges in only 100 iterations on the 100-meter fiber channel, 66% faster than MMDN. Moreover, increased error-correction redundancy delays the error accumulation threshold, effectively enhancing prediction accuracy and outperforming traditional static models(see Supplementary).

Future breakthroughs should focus on more robust error correction strategies: dynamic adaptive coding can flexibly adjust RS code redundancy by real-time monitoring of channel stability metrics such as SSIM fluctuation, balancing bandwidth and reliability; hierarchical error correction frameworks can combine byte-level coarse correction at the physical layer RS code with pixel-level fine optimization using attention networks at the algorithm layer, enhancing local features in mode-coupling sensitive regions and correcting prediction errors; for nonlinear effects in kilometer-scale fibers, further integration of information-theoretic decoding principles with neural networks is needed, employing bidirectional recurrent architectures to truncate error propagation paths; all while maintaining effective data throughput and continuously improving symbol accuracy.

The successful validation of the RRSDN framework demonstrates that the cross-domain integration of information theory and deep learning establishes a new paradigm for information transmission in complex media channels. Its core value lies not only in overcoming the distance-accuracy trade-off in MMF but also in constructing a compatible architecture that integrates discrete coding with continuous network training. These advances will facilitate the transition of MMF technology from laboratory validation to industrial applications, providing critical technical support for extreme scenarios such as endoscopic imaging, deep-sea communication, and deep-space exploration.

**Funding.** This work was supported by National Natural Science Foundation of China 62475175, Chengdu Municipal Bureau of Science and Technology Innovation and R&D Projects 2024-YF05-00268-SN.

**Disclosures.** The author declares no conflicts of interest.

**Data availability.** Data underlying the results presented in this paper are not publicly available at this time but may be obtained from the authors upon reasonable request.

**References.**

[1]Saffman M, Anderson D Z. Mode multiplexing and holographic demultiplexing communication channels on a multimode Fiber [J]. Optics Letters, 1991, 16(5): 300.

[2]Freund R E, Bunge C-A, Ledentsov N N, et al. High-Speed Transmission in Multimode Fibers [J]. Journal of Lightwave Technology, 2010, 28(4): 569–586.

[3]Zhang R, Li X, He Y, et al. Ultra-High Bandwidth Density and Power Efficiency Chip-To-Chip Multimode Transmission through a Rectangular Core Few-Mode Fiber [J]. Laser & Photonics Reviews, 2023, 17(11).

[4]Fan M, Liu K, Zhu J, et al. Flexible single multimode fiber imaging using white LED [J]. Laser Physics, 2025, 35(5): 055102.

[5]Fan W, Chen Z, Yakovlev V V, et al. High-Fidelity Image Reconstruction through Multimode Fiber via Polarization-Enhanced Parametric Speckle Imaging [J]. Laser & Photonics Reviews, 2021, 15(5).

[6]Plöschner M, Tyc T, Čižmár T. Seeing through chaos in multimode Fibres [J]. Nature Photonics, 2015, 9(8): 529–535.

[7]N'Gom M, Norris T B, Michielssen E, et al. Mode control in a multimode fiber through acquiring its transmission matrix from a Reference-less optical System [J]. Optics Letters, 2018, 43(3): 419.

[8]Deng L, Yan J D, Elson D S, et al. Characterization of an imaging multimode optical fiber using a digital Micro-mirror device based single-beam System [J]. Optics Express, 2018, 26(14): 18436.

[9]Chapalo I, Stylianou A, Mégret P, et al. Advances in Optical Fiber Speckle Sensing: A Comprehensive Review [J]. Photonics, 2024, 11(4): 299.

[10]Chamorro-Posada P. TLM Analysis of Multimode Interference Devices [J]. Fiber and Integrated Optics, 2006, 25(1): 1–10.

[11]Fan R, Li L, Zheng Y. High-resolution single-pixel imaging based on a probe of single-mode fiber and hybrid multimode Fiber [J]. Optics & Laser Technology, 2024, 175: 110732.

[12]Zhong J, Wen Z, Li Q, et al. Efficient reference-less transmission matrix retrieval for a multimode fiber using fast Fourier transform [J]. arXiv, 2023.

[13]Wen Z, Dong Z, Deng Q, et al. Single multimode fibre for in vivo Light-field-encoded endoscopic Imaging [J]. Nature Photonics, 2023, 17(8): 679–687.

[14]Schönlieb C-B, Shumaylov Z. Data-driven approaches to inverse problems [Z]. arXiv, 2025(2025).

[15]Karanov B, Chagnon M, Thouin F, et al. End-to-end Deep Learning of Optical Fiber Communications [J]. arXiv, 2018.

[16]Tang P, Zheng K, Yuan W, et al. Learning to transmit images through optical speckle of a multimode fiber with high Fidelity [J]. Applied Physics Letters, 2022, 121(8).

[17]Xiong W, Redding B, Gertler S, et al. Deep learning of ultrafast pulses with a multimode Fiber [J]. APL Photonics, 2020, 5(9).

[18]Wang X, Wang Z, Luo S, et al. Upconversion imaging through multimode fibers based on deep Learning [J]. Optik, 2022, 264: 169444.

[19]Rahmani B, Loterie D, Konstantinou G, et al. Multimode optical fiber transmission with a deep learning Network [J]. Light: Science & Applications, 2018, 7(1).

[20]Resisi S, Popoff S M, Bromberg Y. Image Transmission Through a Dynamically Perturbed Multimode Fiber by Deep Learning [J]. Laser & Photonics Reviews, 2021, 15(10).

[21]Hu S, Liu F, Song B, et al. Multimode fiber image reconstruction based on parallel neural network with small training set under wide temperature Variations [J]. Optics & Laser Technology, 2024, 175: 110815.

[22]Zhang S, Wang Q, Zhou W, et al. Spatial Pilot-aided fast-adapted framework for stable image transmission over long multi-mode Fiber [J]. Optics Express, 2023, 31(23): 37968.

[23]Shaham U, Yamada Y, Negahban S. Understanding adversarial training: Increasing local stability of supervised models through robust Optimization [J]. Neurocomputing, 2018, 307: 195–204.

[24]Huang L, Zhang C, Zhang H. Self-Adaptive Training: Bridging Supervised and Self-Supervised Learning [J]. IEEE Transactions on Pattern Analysis and Machine Intelligence, 2024, 46(3): 1362–1377.

[25]Chen B, Zhang X, Liu S, et al. Self-supervised Scalable Deep Compressed Sensing [J]. International Journal of Computer Vision, 2024, 133(2): 688–723.

[26]Li, Z., Zhou, W., Zhou, Z. et al. Self-supervised dynamic learning for long-term high-fidelity image transmission through unstabilized diffusive media. Nat Commun 15, 1498 (2024).

[27]He K, Zhang X, Ren S, et al. Deep Residual Learning for Image Recognition [Z]. arXiv, 2015(2015).

[28]Li L, Yuan B, Wang Z, et al. Unified architecture for Reed-Solomon decoder combined with burst-error correction[J]. IEEE transactions on very large scale integration (VLSI) systems, 2011, 20(7): 1346-1350.

[29]WANG X, SHI K, WANG J, et al. Error Control Mechanism Based on Reed-Solomon Code for Wireless Networks [Z]. Springer Science and Business Media LLC, 2021(2021–11–23).

[30]Almazmomi N K. Enhanced Reed–Solomon Error Correction Framework for Reliable Data Transmission in Noisy Communication Systems [J]. Internet Technology Letters, 2025.

[31]Liu T, Chen X. Attention-based neural joint source-channel coding of text for point to point and broadcast Channel [J]. Artificial Intelligence Review, 2021, 55(3): 2379–2407.

[32][1]Xia F, Kim K, Eliezer Y, et al. Nonlinear optical encoding enabled by recurrent linear Scattering [J]. Nature Photonics, 2024, 18(10): 1067–1075.